\documentclass{aa}  
\usepackage{graphicx}
\usepackage{txfonts}
\def\draftversion{1} 
\usepackage{amssymb,latexsym,graphicx,natbib,eufrak,times,amsmath,xspace,ifthen}

\ifthenelse{\equal{\draftversion}{1}}{
  \usepackage{xcolor}
  \newcommand{\sep}[1]{\par\begin{color}[rgb]{0,0.4,0}\begin{center}\hrule\end{center}\end{color}\par} 
  \newcommand{\todo}[1]{\begin{color}{red}\ \ifthenelse{\equal{#1}{}} {$\bullet\bullet\bullet$} {$\bullet$\ #1 $\bullet$ \newline}\end{color}} 
  \newcommand{\sk}[1]{\begin{color}[rgb]{0.6,0,0.6}#1\end{color}} 
  \newcommand{\toc}{\par\begin{color}[rgb]{0.6,0,0.6}\begin{center}\hrule\vspace{0.5mm}\begingroup\small\let\cleardoublepage\relax\let\clearpage\relax\mytoc\endgroup\vspace{0.5mm}\hrule\end{center}\end{color}\par} 
  }{
  \newsavebox{\trashcan}

  \newcommand{\sep}[1]{}
  \newcommand{\todo}[1]{}
  
  \newcommand{\sk}[1]{}
  \newcommand{\toc}{}
  }
 \makeatletter\newcommand\mytoc{\@starttoc{toc}}\makeatother 
\long\def\symbolfootnote[#1]#2{\begingroup%
\def\thefootnote{\fnsymbol{footnote}}\footnote[#1]{#2}\endgroup} 

\newcommand{\fig}[2][]{Figure#1~\ref{fig:#2}}

\newcommand{\bb}[1]{\ifmmode \mbox{\boldmath $ #1$} \else  \mbox{\boldmath $#1$} \fi}

\newcommand{\mh}{\ensuremath{\textrm{\,--\,}}}    
\newcommand{\U}[1]{\ensuremath{\mathrm{~#1}}}     
\newcommand{\e}[1]{\ensuremath{\times 10^{#1}}}   

\newcommand{\yr}{\U{yr}}
\newcommand{\Myr}{\U{Myr}}          \newcommand{\myr}{\Myr}
          
\newcommand{\pc}{\U{pc}}

\newcommand{\Msun}{\U{M}_{\odot}}   \newcommand{\msun}{\Msun}
\newcommand{\Msunyr}{\Msun\yr^{-1}} \newcommand{\msunyr}{\Msunyr}
   
\newcommand{\cc}{\U{cm^{-3}}}
\newcommand{\K}{\U{K}}

\newcommand{\dex}{\U{dex}}

\newcommand{\ha}{H$\alpha$\xspace}                         
\newcommand{\aco}{\ensuremath{\alpha_\mathrm{CO}}\xspace}     
\newcommand{\tdep}{\ensuremath{t_\mathrm{dep}}\xspace}        

\newcommand{\lund}{Department of Astronomy and Theoretical Physics, Lund Observatory, Box 43, SE-221 00 Lund, Sweden}

\newcommand{\saclay}{Laboratoire AIM Paris-Saclay, CEA/IRFU/SAp, Universit\'e Paris Diderot, F-91191 Gif-sur-Yvette Cedex, France}

\newcommand{\mpifr}{Max-Planck-Institut f\"ur Radioastronomie (MPIfR), Auf dem H\"ugel 16, 53121 Bonn, Germany}

\begin{document}
\title{Three regimes of CO emission in galaxy mergers}
\titlerunning{Three regimes of CO emission in galaxy mergers}

\author{Florent Renaud\inst{1}, Fr\'ed\'eric Bournaud\inst{2}, Emanuele Daddi\inst{2} \and Axel Wei{\ss}\inst{3}}
\authorrunning{Renaud F., Bournaud F., Daddi E. \& Wei{\ss} A.}
\institute{\lund\\\email{florent@astro.lu.se}\and
\saclay\and
\mpifr}
\authorrunning{Renaud et al.}

\date{Received October 8, 2018; accepted November 15, 2018}

\abstract{The conversion factor \aco from the observable CO(1-0) luminosity to the mass of molecular gas is known to vary between isolated galaxies and some mergers, but the underlying reasons are not clearly understood. Thus, the value(s) of \aco to be adopted remain highly uncertain. To provide better constraints, we apply the large velocity gradient method to a series of hydrodynamical simulations of galaxies and derive the evolution of \aco. We report significant variations of \aco, and identify three distinct regimes: disk galaxies, starbursts and post-burst phases. We show that estimating the star formation rate over $20 \myr$ smooths out some of these differences, but still maintains a distinction between disks and starbursts. We find a tighter correlation of \aco with the gas depletion time than with star formation rate, yet with deviations induced by the transitions to and from the starburst episodes. We conclude that \aco fluctuates because of both the feedback energy and the velocity dispersion. Identifying the phase of an interaction by classical means (e.g. morphology, luminosity) could then help selecting the relevant conversion factor to be used and get more accurate estimates of the molecular masses of galaxies.}
\maketitle

\section{Introduction}

The physical properties of the dense phase of the interstellar medium (ISM) are often estimated using number of tracers (CO, HCH, HNC, HCO$^+$, see e.g. \citealt{Gao2004, Talbi1996, Gracia2006}), each probing different density and temperature ranges. However, to convert the observed luminosities into dynamical quantities like the mass of molecular gas, one must rely on conversion factors like the ratio of molecular gas mass to CO luminosity (\aco, expressed in $\msun \K^{-1} \U{km}^{-1} \U{s} \pc^{-2}$ in the rest of the paper) which have been shown to significantly vary with the galactic environment \citep{Bolatto2013}. Uncertainties are particularly important in interacting galaxies (see e.g. \citealt{Gracia2008}). \citet{Zhu2003} found empirically CO-to-dense gas conversion factors lower than average in actively star forming regions \citep[see also][]{Gracia2008, Sliwa2012, Genzel2015}, while they remain similar to the Milky Way value in high redshift disks \citep{Daddi2010}. The physical reason(s) for these variations are not fully identified, which further prevents the derivation of models. In particular, it is still poorly known whether all major mergers, or only the most extreme ones, have low \aco's.

Numerical simulations and (semi-)analytical models have been used to derive the CO emission and provide estimates of the variations of \aco under different physical conditions. Cloud-scale studies have highlighted the importance of metallicity, dust, turbulence and the local SFR in changing \aco \citep[e.g.][]{Glover2011, Shetty2011, Clark2015, Seifried2017}. In parallel, galaxy-scale works focused on the role of kpc-scale (hydro)dynamics in shaping the dense gas regions and the corresponding \aco's \citep[e.g.][]{Narayanan2011, Narayanan2012, Feldmann2012, Narayanan2014, Bournaud2015, Vollmer2017, Gong2018, Kamenetzky2018}. Such studies typically conduct statistical analyses over a sample of galaxies, but must compromise on the resolution, typically $50 \mh 100 \pc$, i.e. not resolving the cold and dense phase of the ISM. This then calls for sub-grid recipes (either live or in post-processing) to describe the molecular gas and its turbulence, calibrated on Milky Way-like galaxies. However, both the temperature and the turbulence are critical in setting the intrinsic CO emission and the optical depth of the surrounding medium. These models thus might not describe the structure and properties of the star forming sites ($\lesssim 10 \pc$) accurately enough to capture the variations of \aco over a diversity of galaxies.

Using simulations of galaxies with high SFRs ($\approx 50 \Msunyr$, local mergers in their starburst phase and high redshift galaxies), \citet{Bournaud2015} reported a regime of low \aco ($\approx 2$) in starbursting mergers and significantly higher ($\approx 4$) is disks. Here, we examine the time evolution of an interacting and merging galactic pair that experiences starburst phases, along the entire course of its interaction, and explore the transitions between these regimes, using isolated galaxies as references. We conduct our study at parsec-scale resolution, thus explicitly resolving the high densities and cold temperatures ($\sim 10^6 \cc,\, 10 \U{K}$) of molecular clouds and their inner turbulent structure, without having to resort to sub-grid post-processing techniques. However, this comes at the price of a small sample size, rather than a statistical analysis. In this paper, we identify three regimes of \aco along the course of the interaction, and provide fits for these regimes as functions of observable quantities.

\section{Method}

The analysis is performed on a hydrodynamical simulation of the Antennae merger presented in \citet{Renaud2015}, at maximum resolution, i.e. $1.5 \pc$ in the densest regions. Our simulation (and other comparable works, \citealt{Karl2010, Teyssier2010}) predicts two pericenter passages (with a separation of the progenitor galaxies in between), and a third encounter leading to final coalescence. The simulation treats all the ISM at solar metallicity and includes heating from UV background, atomic and molecular cooling, star formation, and stellar feedback in the form of photo-ionization, radiative pressure and type-II supernovae. Supermassive black holes and active galactic nucleus feedback are not implemented. 

We perform a large velocity gradient (LVG) analysis to model CO emission, as presented in \citet{Bournaud2015}. In short, the intrinsic emission is estimated based on the gas density and the temperature in each cells of the simulation, and referring to lookup emission data \citep{Weiss2005}. Then, if the velocity difference between the source and a gas element along the line-of-sight is smaller than the intrinsic width of the emitted line, the flux is absorbed. Conversely, if the velocity shift is large enough, the emission flux is unaffected. This process is repeated for all the grid cells over all columns along the line-of-sight to compute the total CO flux. To compute the CO luminosity to molecular gas mass conversion factor \aco, we consider the mass of the gas denser than $50 \cc$. (Adopting a threshold of $10 \cc$ would induce variations of $10\mh 20 \%$ in \aco, with no systematic trend. This is of the order of the dispersions of our \aco values, as explained below, and does not affect our conclusions.) This criterion gives a mass of molecular gas of $0.2\e{9} \Msun$ in each of the isolated disks, and a decline from $1.9\e{9}$ to $1.3\e{9} \Msun$ for the pair along the interaction due to the enhanced fragmentation of clouds (see also the figure 2 of \citealt{Renaud2014b}). Our method gives $\aco = 4.6$ for a Milky Way-like galaxy \citep{Bournaud2015}, i.e. a value well within the uncertainties of that usually adopted observationally: $\aco = 4.3$ ($\pm 0.1 \dex$, \citealt{Bolatto2013}). To estimate the dispersion of the measured quantities, we tilt the system by $\pm 15^\circ$ along two axes, and re-run the LVG analysis. The dispersion is evaluated as the root mean square of these five lines-of-sight.

\section{Results and interpretation}

\begin{figure}
\centering
\includegraphics{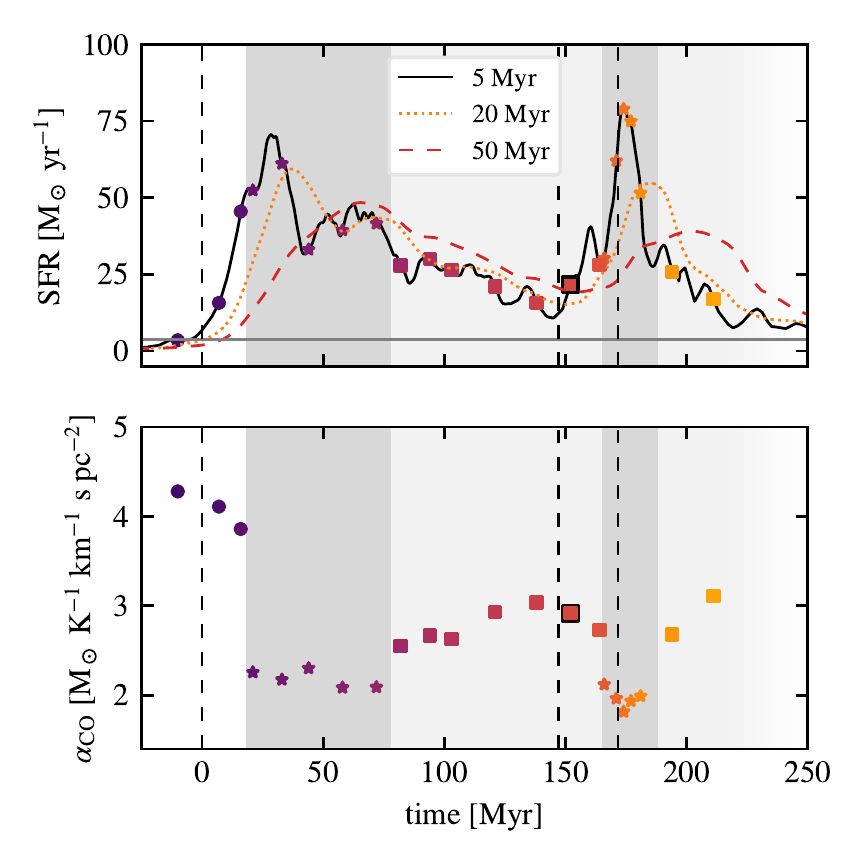}
\caption{Top: evolution of the SFR measured over the last 5, 20 and 50 Myr. Vertical lines indicate the pericenter passages. Symbols mark the instants selected for analysis, with the their color coding time. Their shapes and the shaded areas correspond to the regimes identified below: white, dark gray and light gray for the disk, starburst and post-burst regimes respectively. The black square indicates the instant of best morphological match with the observed Antennae. The gray line is the SFR of the progenitor galaxies, run in isolation (i.e. an almost constant SFR of $\approx 1.5 \msunyr$ for each galaxy). Bottom: evolution of \aco.}
\label{fig:sfr}
\end{figure}

\begin{figure}
\centering
\includegraphics{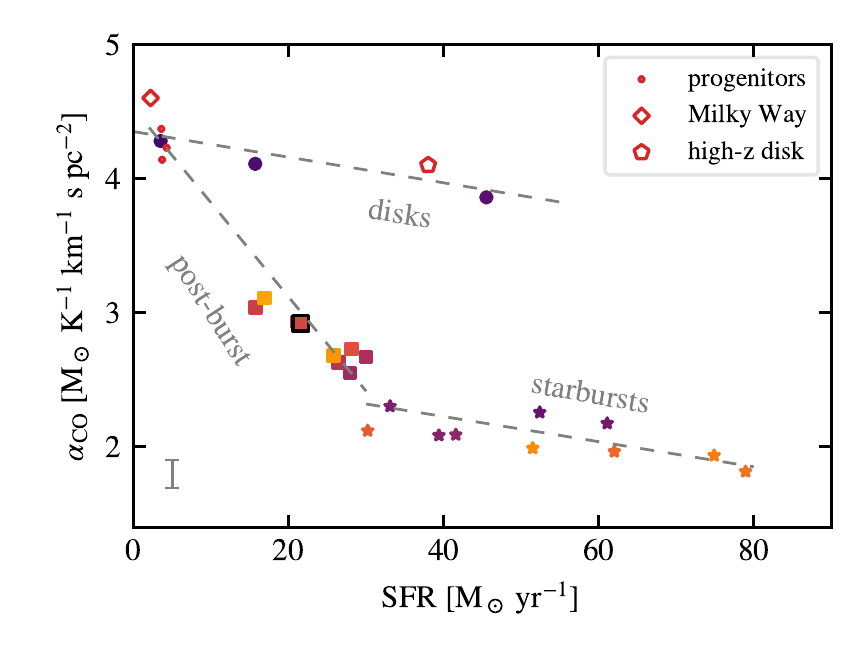}
\caption{\aco as a function of the SFR (measured over $5 \Myr$) in the merger and in simulations of other galaxies. (The system starts in the top-left corner and, at first order, moves clockwise). The symbols are as in \fig{sfr}. The dashed lines indicate linear fits ($\aco = a\, {\rm SFR} + b$) to three regimes: disk galaxies, starbursts, and post-burst phases (see text). The typical dispersions of \aco (estimated by varying the line-of-sight, see text) is shown in the bottom-left corner.}
\label{fig:acosfr}
\end{figure}

\fig{sfr} shows the evolution of the SFR and \aco along the interaction. The SFR is measured over 5, 20 and $50 \Myr$, corresponding to the instantaneous SFR and that from observational tracers of different timescales like \ha and far infrared. \fig{acosfr} shows the relation between \aco and the instantaneous SFR. Additional points corresponds to the progenitor galaxies run in isolation, and to the median values of \aco from simulations of a Milky Way-like galaxy and a $z\sim 2$ gas-rich clumpy disk (both in isolation, see \citealt{Bournaud2015}). In this diagram, we identify three regimes:
\begin{itemize}
\item Disk regime: high \aco, regardless of the SFR. This corresponds to the isolated disks (the Milky Way, the high redshift disk, and our Antennae progenitors), and the earliest instants of the interaction (our first three points), including when the SFR is rising but is already enhanced ($20\mh 40 \Msunyr$).
\item Starburst regime: low \aco, high SFR. These points corresponds to the interacting system, during the peaks of star formation. 
\item Post-burst regime: intermediate \aco, intermediate SFR. This sample gathers exclusively the snapshots found during the decline of the SFR after the first burst and after final coalescence, but with no clear distinction between them.
\end{itemize}
At the moment of best match with the observations, our simulation gives $\aco = 2.9$.

Our model of the Antennae starts in the disk regime and remains there $\sim 20 \Myr$ after the first pericenter passage, i.e. $~25 \Myr$ after the earliest rise of the SFR. Our sample actually comprises a point at $40 \Msunyr$, but still with a \aco comparable to the high-redshift disk. The first reason for this delay in reaching the starburst regime is likely the non-instantaneous propagation of the triggering mechanism leading to the burst. When the progenitors first interact, only a fraction of their gas disks is immediately affected by the boost of star formation. It takes several Myr for the enhancement mechanism to propagate across the disks, depending on the orbital configuration of the encounter. The second reason is related to the local modification of the ISM properties (density, temperature, velocity field) by stellar feedback over timescales of $\sim 1\mh 10 \Myr$. Before enhanced star formation spans the majority of the disks' volume, both these phenomena are local and only affect a fraction of the CO-emitting dense gas.

This is similar to the situation found in our high redshift clumpy disk, where the star formation activity is restricted to an handful of massive clumps. In the interaction however, such a compact and sparse configuration ceases when a larger fraction of the ISM gets compressed by turbulence to high densities \citep{Renaud2014b}, resulting in wide emission lines and shallower density contrasts between the star forming and non-star forming regions than in isolated disks (at low and high redshifts, see \citealt{Bournaud2015}). The boost of star formation then covers larger volumes such that the system moves to the starburst regime. The reason for which the transition between the two regimes takes place in only a few Myr is however not clear. 

Once in the starburst mode, the system moves back and forth between the burst and post-burst regimes, without reaching the disk regime again. This evolution is a complex interplay of tidal disruption, debris falling back onto the disks, global decrease of the SFR, cooling of the feedback bubbles and other mechanisms that drive the system toward lower SFRs and higher \aco's, and back when the late galactic encounters happen. The cost of the simulation prevented us to run it for a long time after coalescence, but it is very likely that the merger remnant will go back to the isolation phase (via the post-burst regime) to reach a low SFR ($\lesssim 10 \Msunyr$) and high \aco ($\approx 4\mh 5$).

If interpreting the enhancement of SFR as an additive process to the pre-interaction activity, one seeks linear relations of the form $\aco = a\, {\rm SFR} + b$. The parameters of the best fits of these relations (plotted in \fig{acosfr}) and the standard deviation $\sigma$ are (for \aco and the SFR expressed in $\msun \K^{-1} \U{km}^{-1} \U{s} \pc^{-2}$ and $\msunyr$):
\begin{equation}
\label{eqn:fit}
\begin{array}{llll}
\textrm{disk regime:} & a = -9.5\e{-3} & b = 4.35 & (\sigma = 0.13)\\
\textrm{starburst regime:} & a = -9.3\e{-3} & b = 2.60 & (\sigma = 0.15)\\
\textrm{post-burst regime:} & a = -70.2\e{-3}& b= 4.52 & (\sigma = 0.26).
\end{array}\nonumber
\end{equation}

\begin{figure}
\centering
\includegraphics{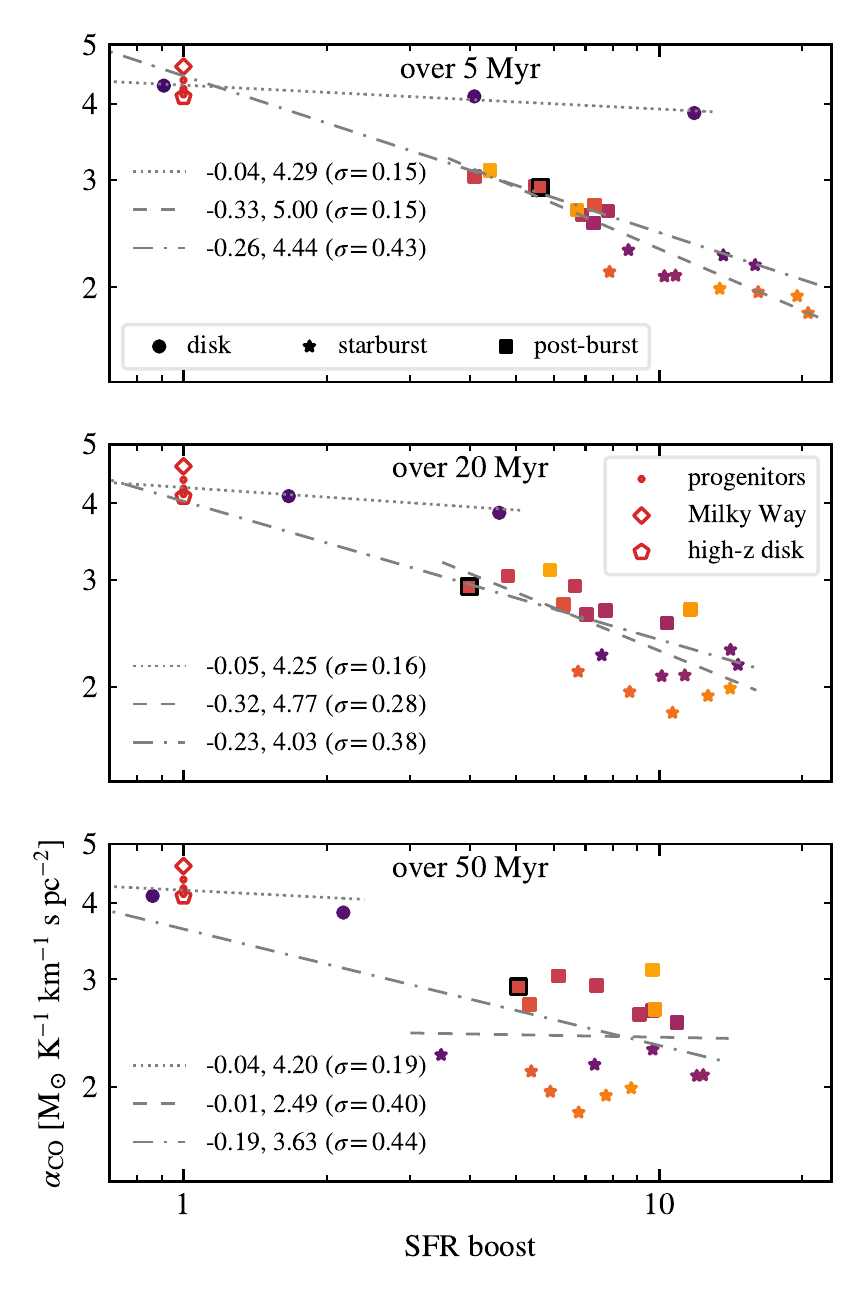}
\caption{\aco as a function of the SFR measured over the last 5 (top), 20 (middle) and 50 Myr (bottom), and normalized to its value in the isolated galaxies. The symbols are as in \fig{sfr}. Dotted, dashed, and dash-dotted lines are fits of the high and low \aco regimes (without distinctions between starburst and post-burst points), and the entire sample respectively. The fitted relation is $\aco = d\, {\rm SFR}^c_{\rm boost}$, with the values $c$ and $d$ indicated in the legend, with the standard deviations ($\sigma$).}
\label{fig:acodelay}
\end{figure}

However, the enhancement of star formation can also be seen as a multiplicative boost of the SFR with respect to that expected for main sequence galaxies. In that case, it is more natural to use power-law relations $\aco = d\, {\rm SFR_{\rm boost}}^c$ to fit our data points, as shown in \fig{acodelay}. Because of the reaction time between the variations of SFR and that of \aco (\fig{sfr}), measuring the SFR over long timescales (e.g. in UV or IR) smooths out its peaks and thus blurs the differences between the starbursts and post-burst regimes. When considering all phases of the interaction (dashed-dotted lines in \fig{acodelay}), the SFR measured over the last $20 \Myr$ provides the least dispersed relation to \aco, corresponding to the timescale of the propagation of feedback to galactic scales.

This relation yields several modes when the SFR boost is measured over timescales shorter than its variations ($\lesssim 50 \Myr$). However, when smoothing it out over longer timescales, the different regimes blend together towards a unimodal relation (but with a larger scatter), in qualitative agreement with that of \citet[their figure 15]{Sargent2014}.

\begin{figure}
\centering
\includegraphics{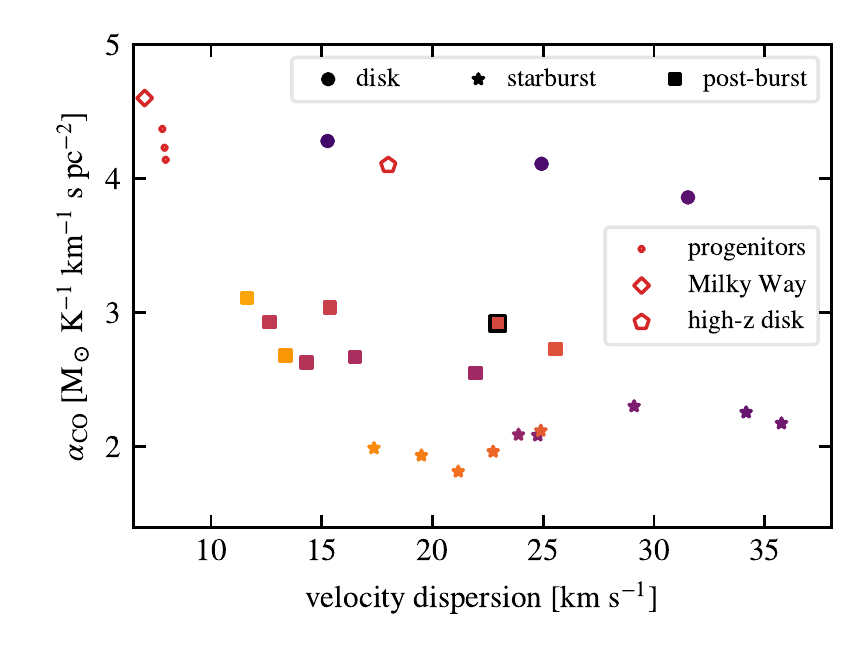}
\caption{\aco as a function of the velocity dispersion of the system. The symbols are as in \fig{sfr}. There is no unequivocal relation(s) between \aco and the velocity dispersion.}
\label{fig:vdisp}
\end{figure}

The decrease of \aco in starbursting galaxies can be interpreted as resulting from the observed increase in velocity dispersion of the ISM \citep{Irwin1994, Ueda2012} which would lower the absorption along the line-of-sigh \citep[e.g.][]{Narayanan2011}. \fig{vdisp} shows however a more complex picture, as no correlation exists between \aco and the velocity dispersion (measured here at the scale of $40 \pc$ and mass-weight averaged over the entire system). A given velocity dispersion can correspond to a wide range of \aco, and no simple relation can be found, even if splitting the points into different regimes. The reason for this is the dual origin of increased velocity dispersion: turbulence and feedback. Indeed, the velocity dispersion is enhanced both before and after a peak of SFR, respectively by turbulence (that triggers the burst) and the feedback that results from it \citep{Renaud2014b}. However, these epochs of enhanced dispersion correspond to different physical states and, as shown in \fig{acosfr}, to different \aco's. Therefore \aco is not a sole function of the velocity dispersion.

This argument also suggests a tighter link with feedback, as already hinted by the delay between the increase of SFR and the drop of \aco noted above. One can consider the SFR normalized by the dense gas mass, i.e. the inverse of the depletion time \tdep, as a proxy for the specific energy injected by stellar feedback. Observations report significantly shorter depletion times in interacting systems than in disks \citep[e.g.][]{Daddi2010b}, but these results suffer from uncertainties on \aco. Yet, simulations do reproduce these differences, without having to choose any \aco. These differences are interpreted as a change of the nature of turbulence inducing efficient compression of the gas \citep{Renaud2012, Renaud2014b, Kraljic2014}. Accounting for the depletion time could thus help tightening the models of \aco by bringing together disks and mergers.

\begin{figure}
\centering
\includegraphics{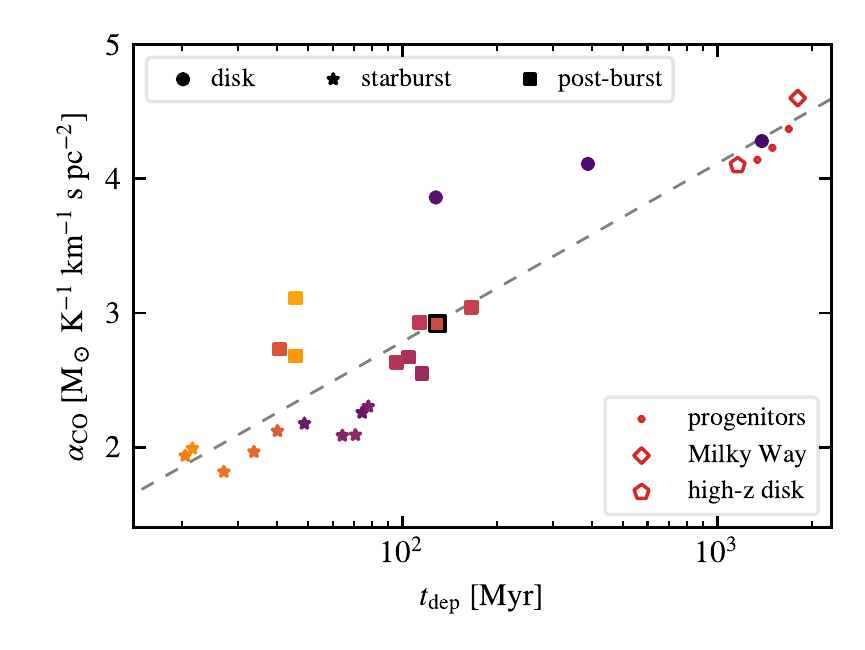}
\caption{\aco as a function of the depletion time from our simulation sample. The symbols are as in \fig{sfr}. The dashed line is the fit $\aco = 1.33 \log{(\tdep / \Myr)} + 0.13$ (with a standard deviation of 0.34), in the units indicated in the figure.}
\label{fig:acotdep}
\end{figure}

\fig{acotdep} shows the relation between $\tdep$ and $\aco$, still with the presence of several modes. The post-burst regime highlighted above ($2.5 \lesssim \aco \lesssim 3.5$) is split into two branches of approximately constant \tdep: that at $\tdep \approx 100 \Myr$ corresponds to the separation phase, after the starburst induced by the first passage but before the second passage (see \fig{sfr}), while that at $\tdep \approx 40 \Myr$ is found after final coalescence. The degeneracy in the \aco \mh \tdep relation along each of these branches illustrates that \aco is not a sole function of the feedback injected in the ISM \citep[see also][]{Bournaud2015}.

It further shows that successive encounters do not induce the same effects (see e.g. \citealt{Renaud2014b} for an Antennae-like system, and \citealt{Renaud2018} for a Cartwheel-like galaxy). In the separation phase, only distant, remote triggers like tidal compression can enhance the SFR, over large scales but at a moderate efficiency, i.e. a mildly enhanced SFR and intermediate $\tdep$. At coalescence however, nuclear inflows induced by tidal torques are the main trigger of star formation. This local and strong activity efficiently fuels gas and convert it into stars, but over a small volume, which leads to a comparably mild SFR but a shorter $\tdep$ than before. While these differences are visible in \fig{acotdep}, \aco is found to better correlate with \tdep than the SFR.

The timescale for the remnant to reach the disk regime again is uncertain. Lower resolution simulations have shown that the period of reduced \tdep lasts several $100 \Myr$ after the last starburst, even though the SFR recovers its pre-interaction value within a few $10 \Myr$ only (see \citealt{Kraljic2014}). The reason for this is still under investigation and it is possible that a similar effect applies to \aco. But such low resolution simulations do not capture the diversity of turbulence in molecular clouds, and are thus biased toward universal values of \tdep and \aco such that they cannot be used to answer this point.

\section{Discussion and conclusions}

By applying the LVG method to hydrodynamical simulations resolving molecular clouds, we derive the evolution of the \aco parameter along a galactic interaction. Our results confirm the existence of distinct \aco's in disks and mergers, even for comparable SFRs \citep[see e.g.][]{Bournaud2015}. By following a galactic system along the course of its interaction and probing the full variation of its SFR, we further identify a third regime corresponding to the post-burst phase, which can last at least $\approx 100 \Myr$. The transitions between the regimes are however very rapid ($\sim 2\mh 10 \Myr$). Our main conclusions are as follows.
\begin{itemize}
\item The variations of \aco are not a smooth, unimodal function of the SFR, because of the modification of the structure and properties of the ISM induced by the interaction, and by star formation and the associated feedback. This leads to three distinct regimes.
\item The first regime gathers galaxies for which no interaction has significantly modified the ISM properties over large scales. It thus comprises isolated galaxies and the earliest phases of interactions (while the SFR is still rising, but can already be significantly enhanced).
\item The starburst regime represents the phases of most active star formation activity, typically $\approx 20-50 \Myr$ after a close passage, and during coalescence. During this regime, the compression of gas triggers star formation over large volumes, from which the feedback induces a drop of \aco.
\item The post-burst regime is found during the separation phase of the galaxies (i.e. between the encounters) and after the final coalescence, while the SFR decreases. \aco increases again, toward the disk regime.
\item The transitions from one regime to the next are set by large-scale hydrodynamics effects convoyed to smaller scales by turbulence. The system crosses the starburst regime in a couple of $10 \Myr$, while it can stays several $100 \Myr$ in the post-starburst mode. The timescale for the transition is of the order of that of the propagation of stellar feedback which alters the structure of the ISM.
\item As a consequence, the latter two regimes are blended together when estimating the SFR over longer timescales, which results in estimations over $20 \Myr$ providing the tightest \aco -SFR relation. Yet, the disks and the earliest phases of the interaction still stand out as a distinct mode.
\item The depletion time, which varies with the nature of turbulence setting the structure of the ISM, provides a better correlation with \aco than the SFR.
\item \aco is not only set by the velocity dispersion or the feedback energy, but is a function of both.
\end{itemize}

Previous works advocated for a smooth transition of \aco with galaxy properties \citep{Narayanan2012, Bolatto2013, French2015}. Our results reveal indeed a more complex picture than a bi-modal relation. The post-burst regime we highlight is the signature of this smooth transition. However, we note that the initial transition from the disk to the starburst regime is extremely fast ($\sim 2 \Myr$), likely because of the interplay of processes acting on short time-scales (dynamics of the interaction, increase of the SFR, feedback) but with the details yet to be identified and understood. This could thus been detected and interpreted as a non-smooth or even discontinuous transition, favoring the idea of bi-modality, while in reality any other transition between regimes is smoother and occurs over longer timescales.

Observational derivations of molecular gas masses based on a universal or bi-modal \aco are thus likely to be affected by important errors. By using the relations we provide here on observational data and accounting for the gas fraction and the merger phase rather than the sole SFR or velocity dispersion, one would likely accentuate the empirical differences in ISM properties and star formation activities in different environments. In the case of post-starburst galaxies for instance, our results suggest to adopt intermediate values of \aco, for at least several $10 \Myr$ after each burst, and in particular after final coalescence. In their sample of post-starburst galaxies, \citet{French2015} measured low SFRs ($\sim 0.1 \msun\yr$, i.e. even smaller values than those probed by our simulations\footnote{The difference could originate from the fact that their galaxies are observed much later after coalescence than our, and/or because of the absence of active galactic nucleus (AGN) feedback in our model what would participate in quenching star formation in our simulation.}), which suggests to adopt values of \aco comparable or even higher to that of disks. Our findings thus support the high values of \aco they used ($\sim 4\mh 6$, instead of choosing a ULIRG value of $\aco = 0.8$, see their figure 12), and thus their conclusion of star formation efficiencies lower in their post-starburst galaxies than in disks despites large amount of CO gas, for a reason yet to be determined (\citealt{French2018}, but see also \citealt{Alatalo2016} for the opposite conclusion from different galaxy selection criteria).

Our finding of several regimes of \aco complements the work of \citet{Narayanan2011} and \citet{Narayanan2014} who adopted a universal sub-grid model to describe turbulence and advocated for a smooth, unimodal transition of the CO-to-H$_2$ factors between disks and mergers. The distinct regimes we find likely originates from the diversity of turbulence our simulations capture. For instance, these variations explain that a given SFR surface density can be found in very different media (e.g. a low redshift merger and a high redshift disk), due to the variation of the depletion time induced by compressive turbulence \citep{Renaud2014b}. Thus, our analysis tells apart galaxies based on the physical trigger of star formation, rather than on the star formation activity itself. As mentioned in \citet{Bournaud2015}, the structure of the ISM, including the non-star forming material, and in particular its clumpy nature influence the propagation of feedback effects that affect the CO emission (and absorption). Although more investigations are needed, this likely explains that \tdep provides a better correlation with \aco than the mere SFR.

The differences between our regimes demonstrate the need to capture the turbulent cascade setting the structure of the ISM down to its dense, cold phase ($\sim 10\K$, $\sim 1\mh 10 \pc$). However, this implies a high cost that yet forbids to model many interacting systems. Our conclusions are thus only based on the Antennae-like major merger presented here. Yet, while the response of galaxies to interactions strongly depends on orbital and intrinsic parameters, the case presented here goes through all the possible phases of interaction and thus provides a wide sampling of the physical conditions one can expect in mergers. 

Our simulations only include a few aspects of feedback, but it is likely that the other mechanisms affect the density structure and the temperature of the star forming regions and thus the emission CO. Our method however provides a value of \aco for a Milky Way-like galaxy close to the observational estimate, which suggests that including the missing physics would not significantly change our conclusions. In the merger however, the lack of AGN feedback might play a more important role in regulating the emission from the nuclear regions. Similarly, accounting for different coolants, including dust, would alter the temperature of the ISM and thus the excitation of CO.

The variations found in \aco suggest a tight but complex link between the physical conditions shaping the ISM and driving star formation and the CO emission. Understanding these could allow to inverse the analysis and infer the underlying physics from the detection of CO emission lines. For instance, it might be possible to tell apart clumpy disks from mergers using several CO emission lines. Spatial variations of the CO emission, as suggested by \citet{Sandstrom2013}, could provide interesting clues, in particular in mergers (Renaud et al., in preparation).

\begin{acknowledgements}
We thank the referee for a constructive report. FR acknowledges support from the Knut and Alice Wallenberg Foundation. This work was supported by GENCI (allocations A0030402192 and A0050402192) and PRACE (allocation pr86di) resources.
\end{acknowledgements}

\bibliographystyle{aa}

\begin{thebibliography}{0}
\expandafter\ifx\csname natexlab\endcsname\relax\def\natexlab#1{#1}\fi

\end{thebibliography}


\begin{thebibliography}{37}
\expandafter\ifx\csname natexlab\endcsname\relax\def\natexlab#1{#1}\fi

\bibitem[{{Alatalo} {et~al.}(2016){Alatalo}, {Aladro}, {Nyland}, {Aalto},
  {Bitsakis}, {Gallagher}, \& {Lanz}}]{Alatalo2016}
{Alatalo}, K., {Aladro}, R., {Nyland}, K., {et~al.} 2016, \apj, 830, 137

\bibitem[{{Bolatto} {et~al.}(2013){Bolatto}, {Wolfire}, \&
  {Leroy}}]{Bolatto2013}
{Bolatto}, A.~D., {Wolfire}, M., \& {Leroy}, A.~K. 2013, \araa, 51, 207

\bibitem[{{Bournaud} {et~al.}(2015){Bournaud}, {Daddi}, {Wei{\ss}}, {Renaud},
  {Mastropietro}, \& {Teyssier}}]{Bournaud2015}
{Bournaud}, F., {Daddi}, E., {Wei{\ss}}, A., {et~al.} 2015, \aap, 575, A56

\bibitem[{{Clark} \& {Glover}(2015)}]{Clark2015}
{Clark}, P.~C. \& {Glover}, S.~C.~O. 2015, \mnras, 452, 2057

\bibitem[{{Daddi} {et~al.}(2010{\natexlab{a}}){Daddi}, {Bournaud}, {Walter},
  {Dannerbauer}, {Carilli}, {Dickinson}, {Elbaz}, {Morrison}, \& {et
  al.}}]{Daddi2010}
{Daddi}, E., {Bournaud}, F., {Walter}, F., {et~al.} 2010{\natexlab{a}}, \apj,
  713, 686

\bibitem[{{Daddi} {et~al.}(2010{\natexlab{b}}){Daddi}, {Elbaz}, {Walter},
  {Bournaud}, {Salmi}, {Carilli}, {Dannerbauer}, {Dickinson}, \& {et
  al.}}]{Daddi2010b}
{Daddi}, E., {Elbaz}, D., {Walter}, F., {et~al.} 2010{\natexlab{b}}, \apjl,
  714, L118

\bibitem[{{Feldmann} {et~al.}(2012){Feldmann}, {Gnedin}, \&
  {Kravtsov}}]{Feldmann2012}
{Feldmann}, R., {Gnedin}, N.~Y., \& {Kravtsov}, A.~V. 2012, \apj, 758, 127

\bibitem[{{French} {et~al.}(2015){French}, {Yang}, {Zabludoff}, {Narayanan},
  {Shirley}, {Walter}, {Smith}, \& {Tremonti}}]{French2015}
{French}, K.~D., {Yang}, Y., {Zabludoff}, A., {et~al.} 2015, \apj, 801, 1

\bibitem[{{French} {et~al.}(2018){French}, {Zabludoff}, {Yoon}, {Shirley},
  {Yang}, {Smercina}, {Smith}, \& {Narayanan}}]{French2018}
{French}, K.~D., {Zabludoff}, A.~I., {Yoon}, I., {et~al.} 2018, \apj, 861, 123

\bibitem[{{Gao} \& {Solomon}(2004)}]{Gao2004}
{Gao}, Y. \& {Solomon}, P.~M. 2004, \apjs, 152, 63

\bibitem[{{Genzel} {et~al.}(2015){Genzel}, {Tacconi}, {Lutz}, {Saintonge},
  {Berta}, {Magnelli}, {Combes}, {Garc{\'{\i}}a-Burillo}, {Neri}, {Bolatto},
  {Contini}, {Lilly}, {Boissier}, {Boone}, {Bouch{\'e}}, {Bournaud}, {Burkert},
  {Carollo}, {Colina}, {Cooper}, {Cox}, {Feruglio}, {F{\"o}rster Schreiber},
  {Freundlich}, {Gracia-Carpio}, {Juneau}, {Kovac}, {Lippa}, {Naab}, {Salome},
  {Renzini}, {Sternberg}, {Walter}, {Weiner}, {Weiss}, \& {Wuyts}}]{Genzel2015}
{Genzel}, R., {Tacconi}, L.~J., {Lutz}, D., {et~al.} 2015, \apj, 800, 20

\bibitem[{{Glover} \& {Mac Low}(2011)}]{Glover2011}
{Glover}, S.~C.~O. \& {Mac Low}, M.-M. 2011, \mnras, 412, 337

\bibitem[{{Gong} {et~al.}(2018){Gong}, {Ostriker}, \& {Kim}}]{Gong2018}
{Gong}, M., {Ostriker}, E.~C., \& {Kim}, C.-G. 2018, \apj, 858, 16

\bibitem[{{Graci{\'a}-Carpio} {et~al.}(2006){Graci{\'a}-Carpio},
  {Garc{\'{\i}}a-Burillo}, {Planesas}, \& {Colina}}]{Gracia2006}
{Graci{\'a}-Carpio}, J., {Garc{\'{\i}}a-Burillo}, S., {Planesas}, P., \&
  {Colina}, L. 2006, \apjl, 640, L135

\bibitem[{{Graci{\'a}-Carpio} {et~al.}(2008){Graci{\'a}-Carpio},
  {Garc{\'{\i}}a-Burillo}, {Planesas}, {Fuente}, \& {Usero}}]{Gracia2008}
{Graci{\'a}-Carpio}, J., {Garc{\'{\i}}a-Burillo}, S., {Planesas}, P., {Fuente},
  A., \& {Usero}, A. 2008, \aap, 479, 703

\bibitem[{{Irwin}(1994)}]{Irwin1994}
{Irwin}, J.~A. 1994, \apj, 429, 618

\bibitem[{{Kamenetzky} {et~al.}(2018){Kamenetzky}, {Privon}, \&
  {Narayanan}}]{Kamenetzky2018}
{Kamenetzky}, J., {Privon}, G.~C., \& {Narayanan}, D. 2018, \apj, 859, 9

\bibitem[{{Karl} {et~al.}(2010){Karl}, {Naab}, {Johansson}, {Kotarba}, {Boily},
  {Renaud}, \& {Theis}}]{Karl2010}
{Karl}, S.~J., {Naab}, T., {Johansson}, P.~H., {et~al.} 2010, \apjl, 715, L88

\bibitem[{{Kraljic}(2014)}]{Kraljic2014}
{Kraljic}, K. 2014, {Links between galaxy evolution, morphology and internal
  physical processes} (PhD Thesis, Universite Paris 11, 2014)

\bibitem[{{Narayanan} {et~al.}(2011){Narayanan}, {Krumholz}, {Ostriker}, \&
  {Hernquist}}]{Narayanan2011}
{Narayanan}, D., {Krumholz}, M., {Ostriker}, E.~C., \& {Hernquist}, L. 2011,
  \mnras, 418, 664

\bibitem[{{Narayanan} \& {Krumholz}(2014)}]{Narayanan2014}
{Narayanan}, D. \& {Krumholz}, M.~R. 2014, \mnras, 442, 1411

\bibitem[{{Narayanan} {et~al.}(2012){Narayanan}, {Krumholz}, {Ostriker}, \&
  {Hernquist}}]{Narayanan2012}
{Narayanan}, D., {Krumholz}, M.~R., {Ostriker}, E.~C., \& {Hernquist}, L. 2012,
  \mnras, 421, 3127

\bibitem[{{Renaud} {et~al.}(2018){Renaud}, {Athanassoula}, {Amram}, {Bosma},
  {Bournaud}, {Duc}, {Epinat}, {Fensch}, {Kraljic}, {Perret}, \&
  {Struck}}]{Renaud2018}
{Renaud}, F., {Athanassoula}, E., {Amram}, P., {et~al.} 2018, \mnras, 473, 585

\bibitem[{{Renaud} {et~al.}(2015){Renaud}, {Bournaud}, \& {Duc}}]{Renaud2015}
{Renaud}, F., {Bournaud}, F., \& {Duc}, P.-A. 2015, \mnras, 446, 2038

\bibitem[{{Renaud} {et~al.}(2014){Renaud}, {Bournaud}, {Kraljic}, \&
  {Duc}}]{Renaud2014b}
{Renaud}, F., {Bournaud}, F., {Kraljic}, K., \& {Duc}, P.-A. 2014, \mnras, 442,
  L33

\bibitem[{{Renaud} {et~al.}(2012){Renaud}, {Kraljic}, \&
  {Bournaud}}]{Renaud2012}
{Renaud}, F., {Kraljic}, K., \& {Bournaud}, F. 2012, \apjl, 760, L16

\bibitem[{{Sandstrom} {et~al.}(2013){Sandstrom}, {Leroy}, {Walter}, {Bolatto},
  {Croxall}, {Draine}, {Wilson}, {Wolfire}, {Calzetti}, {Kennicutt}, {Aniano},
  {Donovan Meyer}, {Usero}, {Bigiel}, {Brinks}, {de Blok}, {Crocker}, {Dale},
  {Engelbracht}, {Galametz}, {Groves}, {Hunt}, {Koda}, {Kreckel}, {Linz},
  {Meidt}, {Pellegrini}, {Rix}, {Roussel}, {Schinnerer}, {Schruba}, {Schuster},
  {Skibba}, {van der Laan}, {Appleton}, {Armus}, {Brandl}, {Gordon}, {Hinz},
  {Krause}, {Montiel}, {Sauvage}, {Schmiedeke}, {Smith}, \&
  {Vigroux}}]{Sandstrom2013}
{Sandstrom}, K.~M., {Leroy}, A.~K., {Walter}, F., {et~al.} 2013, \apj, 777, 5

\bibitem[{{Sargent} {et~al.}(2014){Sargent}, {Daddi}, {B{\'e}thermin},
  {Aussel}, {Magdis}, {Hwang}, {Juneau}, {Elbaz}, \& {da Cunha}}]{Sargent2014}
{Sargent}, M.~T., {Daddi}, E., {B{\'e}thermin}, M., {et~al.} 2014, \apj, 793,
  19

\bibitem[{{Seifried} {et~al.}(2017){Seifried}, {Walch}, {Girichidis}, {Naab},
  {W{\"u}nsch}, {Klessen}, {Glover}, {Peters}, \& {Clark}}]{Seifried2017}
{Seifried}, D., {Walch}, S., {Girichidis}, P., {et~al.} 2017, \mnras, 472, 4797

\bibitem[{{Shetty} {et~al.}(2011){Shetty}, {Glover}, {Dullemond}, \&
  {Klessen}}]{Shetty2011}
{Shetty}, R., {Glover}, S.~C., {Dullemond}, C.~P., \& {Klessen}, R.~S. 2011,
  \mnras, 412, 1686

\bibitem[{{Sliwa} {et~al.}(2012){Sliwa}, {Wilson}, {Petitpas}, {Armus},
  {Juvela}, {Matsushita}, {Peck}, \& {Yun}}]{Sliwa2012}
{Sliwa}, K., {Wilson}, C.~D., {Petitpas}, G.~R., {et~al.} 2012, \apj, 753, 46

\bibitem[{{Talbi} {et~al.}(1996){Talbi}, {Ellinger}, \& {Herbst}}]{Talbi1996}
{Talbi}, D., {Ellinger}, Y., \& {Herbst}, E. 1996, \aap, 314, 688

\bibitem[{{Teyssier} {et~al.}(2010){Teyssier}, {Chapon}, \&
  {Bournaud}}]{Teyssier2010}
{Teyssier}, R., {Chapon}, D., \& {Bournaud}, F. 2010, \apjl, 720, L149

\bibitem[{{Ueda} {et~al.}(2012){Ueda}, {Iono}, {Petitpas}, {Yun}, {Ho},
  {Kawabe}, {Mao}, {Mart{\'{\i}}n}, \& {et al.}}]{Ueda2012}
{Ueda}, J., {Iono}, D., {Petitpas}, G., {et~al.} 2012, \apj, 745, 65

\bibitem[{{Vollmer} {et~al.}(2017){Vollmer}, {Gratier}, {Braine}, \&
  {Bot}}]{Vollmer2017}
{Vollmer}, B., {Gratier}, P., {Braine}, J., \& {Bot}, C. 2017, \aap, 602, A51

\bibitem[{{Wei{\ss}} {et~al.}(2005){Wei{\ss}}, {Walter}, \&
  {Scoville}}]{Weiss2005}
{Wei{\ss}}, A., {Walter}, F., \& {Scoville}, N.~Z. 2005, \aap, 438, 533

\bibitem[{{Zhu} {et~al.}(2003){Zhu}, {Seaquist}, \& {Kuno}}]{Zhu2003}
{Zhu}, M., {Seaquist}, E.~R., \& {Kuno}, N. 2003, \apj, 588, 243

\end{thebibliography}

\end{document}